\documentclass{aastex631}

\usepackage [english]{babel}
\usepackage [autostyle, english = american]{csquotes}
\MakeOuterQuote{"}
\newcommand{\mgii}{\mbox{Mg\,{\scshape ii}}}
\newcommand{\fesc}{\mbox{$f_{\rm esc}^{\rm LyC}$}}
\newcommand{\oii}{\mbox{[O\,{\scshape ii}}]}
\newcommand{\oiii}{\mbox{[O\,{\scshape iii}}]}
\begin{document}

\title{Using \ion{Mg}{2} Doublet to Predict the Lyman Continuum Escape Fraction from 14 HETDEX Galaxies}

\author[0000-0001-5331-065X]{Victoria Salazar}
\affiliation{Department of Astronomy, The University of Texas at Austin, 2515 Speedway Boulevard, Austin, TX 78712, USA}

\author[0000-0002-6085-5073]{Floriane Leclercq}
\affiliation{Department of Astronomy, The University of Texas at Austin, 2515 Speedway Boulevard, Austin, TX 78712, USA}

\author[0000-0002-0302-2577]{John Chisholm}
\affiliation{Department of Astronomy, The University of Texas at Austin, 2515 Speedway Boulevard, Austin, TX 78712, USA}

\author[0000-0001-6717-7685]{Gary J. Hill}
\affiliation{Department of Astronomy, The University of Texas at Austin, 2515 Speedway Boulevard, Austin, TX 78712, USA}
\affiliation{McDonald Observatory, The University of Texas at Austin, Austin, TX 78712, USA}

\author[0000-0003-2307-0629]{Gregory R. Zeimann}
\affil{Hobby-Eberly Telescope, University of Texas at Austin, Austin, TX 78712, USA}

\begin{abstract}

Indirect diagnostics of Lyman continuum (LyC) escape are needed to constrain which sources reionized the universe.
We used \mgii\ to predict the LyC escape fraction (\fesc) in 14 galaxies selected from the Hobby-Eberly Telescope Dark Energy Experiment solely based upon their \mgii\ properties. Using the Low Resolution Spectrograph on HET, we identified 7 and 5 possible LyC leakers depending on the method, with \fesc\ ranging from 3 to 80$\%$. 
Interestingly, our targets display diverse \oiii/\oii\ ratios (O32), with strong inferred LyC candidates showing lower O32 values than previous confirmed LyC leaker samples. Additionally, a correlation between dust and \fesc\ was identified. Upcoming Hubble Space Telescope/Cosmic Origins Spectrograph LyC observations of our sample will test if \mgii\ and dust are predictors of \fesc, providing insights for future JWST studies of high-redshift galaxies.

\end{abstract}

\section{Introduction} 
\label{sec:intro}

\begin{figure}[ht!]
\plotone{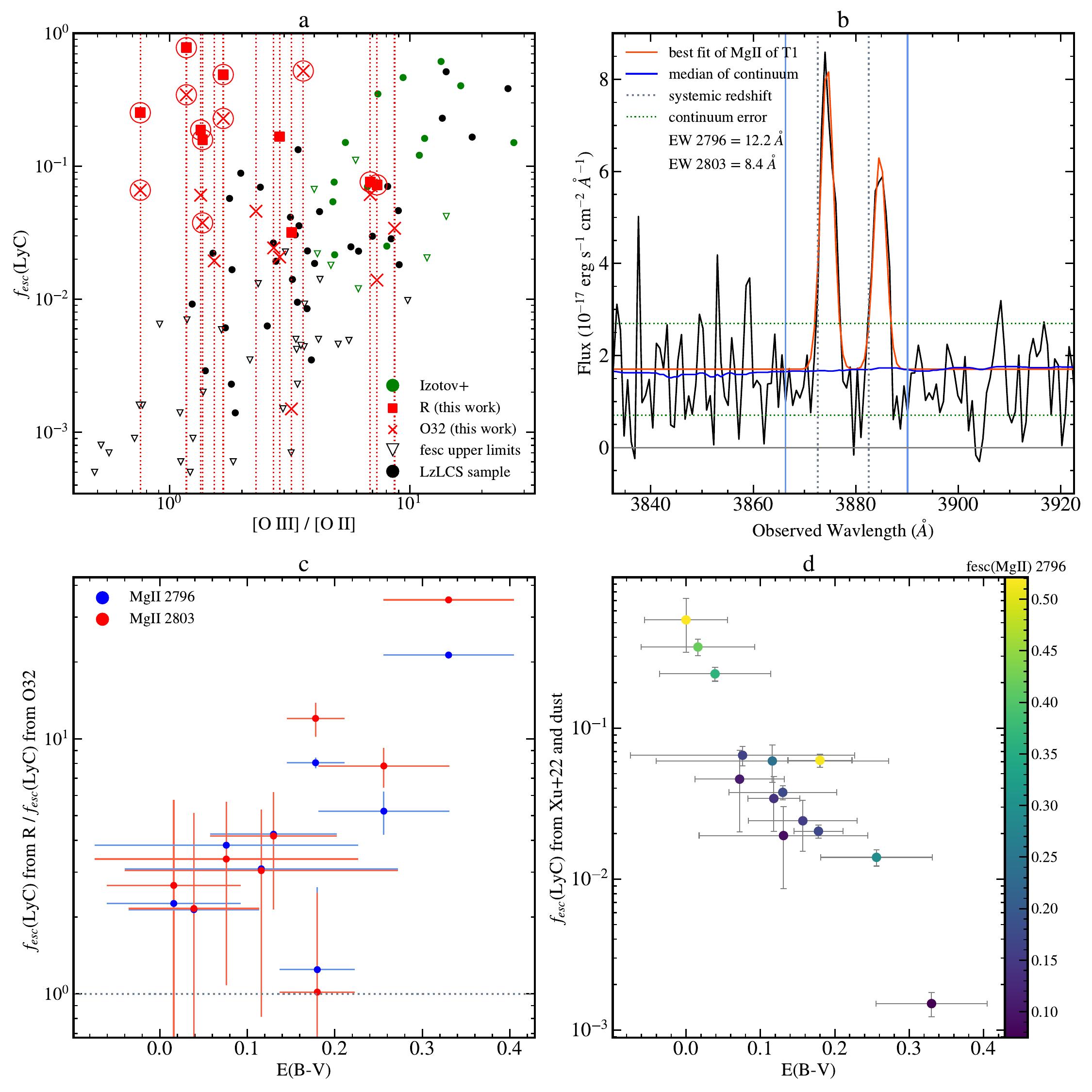}
\caption{(a): Comparison of our sample (red) to literature (black: \citealt{Flury2022}; green: \citealt{Izotov16a, Izotov16b, Izotov18a, Izotov18b, Izotov21}). The red circled points show our LyC leaker candidates from the O32 (crosses) and R (squares) methods (Sects.~\ref{sec:floats} and ~\ref{subsec:lyc}). (b): \mgii\ doublet (black) with best-fit in orange and continuum in blue (Sect~\ref{subsec:spec}). 
(c): The predicted LyC values from the R and O32 methods are compared to the dust values (Sect~\ref{subsec:dust}). (d): Relation between the predicted \fesc\ and dust, with points color coded by the \mgii\ escape fraction. 
\label{fig:general}}
\end{figure}

The epoch of reionization (EoR) is when the gas between galaxies in the early universe became ionized from Lyman continuum (LyC, $\lambda < 912$ \AA) photons emitted by galaxies. The LyC escape fraction (\fesc) is the fraction of the ionizing photons produced by the stars that escape from the galaxy. The neutral gas between galaxies during EoR prevents LyC observations. Therefore, we need indirect techniques calibrated on low-redshift galaxies, where LyC is observable, to infer \fesc\ at high-redshift.
%to better understand the sources of cosmic reionization. 
Previous samples of local leakers and non-leakers (\citealt{Izotov16a, Izotov16b, Izotov18a, Izotov18b, Izotov21, Flury2022}) with diverse \oiii/\oii\ ratios (O32), UV slopes and star formation rate surface densities exhibited correlations, with significant scatter, between \fesc\ and indirect observables like O32 ratios (Figure~\ref{fig:general}a). As a resonant line tracing neutral gas, \mgii\ has recently been proposed as a promising tracer of LyC leakage \citep[hereafter C20]{Henry2018, Chisholm2020}. Using ground spectroscopy, we analyzed the \mgii\ doublet lines from 14 $z\sim0.4$ galaxies selected from their bright \mgii\ and \oii\ emission in the Hobby-Eberly Telescope Dark Energy Experiment survey \citep[Figure~\ref{fig:general}b]{Gebhardt2021, MentuchCooper2023}. The \mgii\ 2796 \AA\ EW values range from 3 to 30 \AA\ with an average of 9.6 \AA, which is higher than, e.g., \citealt{Xu2022} (hereafter X22, average of 5.6 \AA). \\

\section{Methods} 
\label{sec:style}
\subsection{Spectral Measurements}
\label{subsec:spec}

Our targets were observed with Low Resolution Spectrograph (LRS2) \citep{Chonis_2016} installed on HET \citep{Ramsey98, Hill21} for $\approx$25 minutes per target (Leclercq et al. in prep). LRS2 is an integral field fiber-fed spectrograph with coverage from 3650 \AA\ to 10500 \AA\ at resolution R = 1500$-$2500. The data were reduced using the Panacea\footnote{https://github.com/grzeimann/Panacea} pipeline including: fiber extraction, wavelength calibration, astrometry, flux calibration and sky subtraction. The spectra were extracted using a 1.5'' radius aperture to compare to SDSS spectra (when available).
The \mgii\ doublets were fitted with two Gaussian functions. 
The continuum was removed by performing median filtering. The continuum error was estimated using the standard deviation in a 100 \AA\ window avoiding the \mgii\ lines (vertical blue lines in Figure \ref{fig:general}b). 
Using an integrated flux signal to noise ratio threshold of 4, five galaxies show no detection of the \mgii\ 2803 \AA\ line. We used a Monte Carlo approach to measure the errors.

\subsection{Dust Measurements}
\label{subsec:dust}

The $pPXF$ function (\citealt{Cappellari2022}) estimated and removed the continuum and stellar absorption. One Gaussian was fit to the H$\delta$ and H$\gamma$ lines, while two were fit to the H$\alpha$ and H$\beta$ lines to account for significant wings. After correcting the fluxes for the Milky Way extinction (\citealt{Green2018}), we used the $redneb$ function (\citealt{Flury2022}) to estimate the color excess $E(B-V)$ from the Balmer emission lines with fixed gas temperature ($10^4$ K) and density ($10^2$  cm$^{-3}$). We used the SDSS spectrum for one object because the LRS2 data produced a nonphysical $E(B-V)$ value. We note that the $E(B-V)$ values are statistically consistent between the LRS2 and SDSS when data are available.\newline\\

\subsection{\fesc\ Predictions}
\label{subsec:lyc}

We use two methods to predict \fesc. The first method, referred to as the R method, uses dust attenuation and the \mgii\ flux ratio ($R = F_{2796}/F_{2803}$) to estimate the neutral hydrogen column density following C20. The second method, referred to as the O32 method, uses photoionization models following X22. The \oii\ and \oiii\ lines were fitted similarly as \mgii\ (section~\ref{subsec:spec}) with an additional component to capture the broad wings. Their fluxes were dust corrected to find the intrinsic \mgii\ flux and \mgii\ escape fractions assuming a gas-phase metallicity of 10\% solar (X22). 
%The  were obtained by dividing the observed and intrinsic \mgii\ flux. 
The \mgii\ escape fraction was dust attenuated to determine \fesc.\\

\section{Results and Discussion} 
\label{sec:floats}

Using the R and O32 methods, we found 7 and 5 LyC leaker candidates at $>$2$\sigma$ significance (Figure~\ref{fig:general}a), respectively. Our \fesc\ values range from 3 to 80$\%$ (Figure \ref{fig:general}a). If confirmed, our 80$\%$ leakage candidate will be one of the strongest LyC leakers known, providing a novel window into the LyC escape. Our sample includes weak and strong potential leakers with broad O32 ratios range (0.74 to 8.51; Figure~\ref{fig:general}a). Notably, the O32 value of our strongest inferred LyC emitter is one order of magnitude lower than that of previous strong leakers. Figure~\ref{fig:general}c shows that the R method predicts higher \fesc\ values compared to the O32 method, with higher discrepancy at higher dust values. Finally, Figure~\ref{fig:general}d shows a negative correlation between the dust and \fesc\ measured using photoionization models method, suggesting that a lower dust value allows for a higher amount of LyC to escape. This correlation is not as obvious using the R method. The R method contains large errors on weakest LyC values which inhibits firm conclusions, but there is a slight trend in increasing \fesc\ with increasing $R$. %Higher SNR data will be needed to validate the relation with $R$ and \fesc, and confirm that \mgii\ and dust alone are good predictors of LyC.\\

%\section{Conclusion} 

%Using \mgii, we identified 7 and 5 potential LyC leaker candidates within our HETDEX sample using the \cite{Chisholm2020} and \cite{Xu2022} methods, respectively. 
Upcoming Hubble Space Telescope/Cosmic Origins Spectrograph LyC measurements of our targets will test and determine the limits of the methods. Future work will be increase the sample size to better understand if \mgii\ and dust alone can predict \fesc. A strong relation will help JWST uncover the main sources that drove the EoR.

\bibliography{sample631}{}

\bibliographystyle{aasjournal}

\end{document}